\documentclass[showpacs,preprintnumbers,amssymb,twocolumn]{revtex4-2}
\usepackage{graphicx}
\usepackage{dcolumn}
\usepackage{color}
\usepackage{bm}
\usepackage{amsmath}
\usepackage{amssymb}
\usepackage{epsfig}
\usepackage{amsfonts}
\usepackage{lineno,hyperref}
\usepackage{array}
\usepackage{float}
\usepackage{microtype}
\usepackage{multirow}
\usepackage{adjustbox}
\usepackage[english]{babel}
\usepackage{epstopdf}
\usepackage{blindtext}
\usepackage{subcaption}
\usepackage[a4paper, total={7.5in, 10in}]{geometry}

\def \a{\alpha}
\def \b{\beta}

\def \g{\gamma}
\def \d{\delta}
\def \k{\kappa}
\def \s{\sqrt}
\def \be{\begin{equation}}
\def \ee{\end{equation}}
\def \ben{\begin{eqnarray}}
\def \een{\end{eqnarray}}
\def \o{\omega}
\def \O{\Omega}
\def \d{\delta}
\def \sg{\sigma}
\def \p{\partial}
\def \t{\theta}

\def \L{\mathcal{L}_m}
\def \A{\mathcal{A}}
\def \r{\rho}
\def \k{\kappa}

\def \non{\nonumber}
\def \sq{\square}
\def \nb{\nabla}
\def \T{\Theta}

\def \th{\hat{\Theta}}
\def \rf{\tilde{\r}}
\def \pf{\tilde{p}}

\begin{document}
\title{Reconstruction of $f(R,T)$ gravity model via the Raychaudhuri equation}

\author{Arijit Panda}
\email{arijitpanda260195@gmail.com}
\affiliation{Department of Physics, Raiganj University, Raiganj, Uttar Dinajpur, West Bengal, India, 733 134. $\&$\\
Department of Physics, Prabhat Kumar College, Contai, Purba Medinipur, India, 721 404.}

\author{Surajit Das}
\email{surajit.cbpbu20@gmail.com}
\affiliation{Department of Physics, Cooch Behar Panchanan Barma University, Vivekananda Street, Coochbehar, West Bengal, India, 736 101.}

\author{Goutam Manna$^a$}
\altaffiliation{goutammanna.pkc@gmail.com\\$^a$Corresponding author}
\affiliation{Department of Physics, Prabhat Kumar College, Contai, Purba Medinipur 721404, India $\&$\\ Institute of Astronomy Space and Earth Science, Kolkata 700054, India}

\author{Saibal Ray}
\email{saibal.ray@gla.ac.in}
\affiliation{Centre for Cosmology, Astrophysics and Space Science (CCASS), GLA University, Mathura 281406, Uttar Pradesh, India}

\date{\today}

\begin{abstract}
In this work, we investigate for an analytical solution under modified gravity theory, specifically the $f(R,T)$ gravity for two different eras, i.e., matter and dark energy dominated accelerating universe from completely geometrical and mathematical point of view with the help of the Raychaudhuri equation. To construct $f(R,T)$ gravity model, we consider the functional form of $f(R,T)$ as the sum of two independent functions of the Ricci scalar $R$ and the trace of the energy-momentum tensor $T$, respectively. Under the consideration of this type of power law expansion of the Universe we have studied the viability, stability and all the energy conditions. We note that the strong energy condition is not satisfied in our model, which is obvious for the present scenario of the Universe.
\end{abstract}



\maketitle

\section{Introduction}
 With the advancements in technology in the field of astronomy, astrophysics, cosmology and data science, it is now a well-known fact that our universe is expanding at a higher rate than before. Using data from type Ia supernovae (SNe Ia), baryon acoustic oscillations (BAO) and the cosmic microwave background (WMAP7)~\cite{Riess,Perlmutter,Komatsu}, it has also been discovered that there exists an exotic component in our universe known as dark matter~\cite{Bernardis,Hanany,Peebles,Padmanabhan,Clifton}. The mainstream theory of general relativity cannot properly explain why the universe is accelerating. This puts at risk to Einstein's fruitful theory of general relativity (GR) and demands for certain changes into it. 
 
 In recent years, a new approach to solve these challenges has emerged under the heading of the modified theory of gravity. Among those theories, a popular theory is `$f(R)$ theory of gravity'~\cite{Carroll}, in which the action is considered to be an analytic function of Ricci scalar, i.e., $f(R)$, rather than $R$ itself. This function $f(R)$ can also contain the higher orders of $R$. A lot of works have been done time and again under this type of theory~\cite{Starobinsky,Song,Tsu,Nojiri1,Nojiri2,Cognola,Capo1,Wang,Felice,Sotiriou}. A brief review of this field can be found in the paper~\cite{Capo2}. However, Harko et al.~\cite{Harko}, generalized $f(R)$ theory by taking into account an arbitrary coupling between matter and spacetime geometry in the form  $f(R,T)$, where $R$ is the Ricci scalar and $T$ is the trace of the energy-momentum tensor.  

 The primary motivation for the use of the term trace is to take into consideration exotic imperfect fluids and quantum effects. It is a well-known fact that $f(R,T)$ gravity is also capable of describing the late-time cosmic acceleration~\cite{Nagpal1}. Over the years, extensive study has also been conducted in this area. There have been studies on the testing of energy conditions~\cite{Alvarenga}, alternatives to dark matter~\cite{Zaregonbadi}, cosmological non-viability tests~\cite{Velten}, configurations of white dwarfs~\cite{Carvalho}, the Palatini formulation~\cite{Wu} and the genesis of weird stars~\cite{Deb}. Here one may find some excellent works of $f(R,T)$ gravity theory in details~\cite{FG,Sharif1,Sharif2,Houndjo,Jamil1,Yousaf,Phras,Alves,Sharma,Nagpal,Yousaf2,Das,Singh,Nojiri3,Capo3,Nojiri4,Faraoni1,Bamba}. To address the difficulties presented by $f(R,T)$ gravity, some observational tests have been conducted~\cite{Sardar,Moraes,Gamonal}. Therefore, it is now widely acknowledged that the $f(R,T)$ theory of gravity is a successful theory for providing a more accurate and concise description of the universe. 

The Raychaudhuri equation~\cite{Ray55,Ray57,Ray,Kar,Tipler1,Tipler2} plays an important role in the aspects of modern cosmology. The equation simply tells us about the investigation and study of the temporal behaviour of a geodesics congruence, observed by another neighbouring geodesics congruence on its evolution. The evolution essentially characterizes the flow of congruence in a particular spacetime background, and the flow is generated by an integral curve of the tangent vectors (may be timelike or null). So, the equation tells about the rate of change of the expansion of the congruence (which is a coordinate-independent measurement) relates to the gravitational effect. Since the Raychaudhuri equation is completely a geometric identity, known as Codazzi-Raychaudhuri identity~\cite{Frankel,Zafiris,Carter}, it is a fundamental equation to study the exact solutions of Einstein field equation in general relativity~\cite{Dunsby}. The evolution of geodesic congruence could be a case for both, focusing and non-focusing. If strong energy condition is satisfied related to the Ricci tensor, then the corresponding convergence condition implies the focusing of congruence~\cite{Poisson}. On the other hand, one may also shown that conditionally a non-focusing geodesic can happen as a bouncing cosmological scenario in the background of an emergent spacetime~\cite{Surajit}.

Our motivation is to construct a mathematically general form of an analytical solution in $f(R,T)=f_1(R)+f_2(T)$ gravity model~\cite{Harko2011,Mishra2020} for two different epochs from the geometrical point of view. Since the Raychaudhuri equation is completely geometrical in nature, we can utilize it in our case. We have chosen a particular form of cosmological scale factor leading to an eternally accelerating power law expansion of the universe to establish the functional form of $f(R,T)$ gravity model in two different examples such as (i) the matter dominated era, and (ii) the dark energy-dominated accelerating universe era. We have simply used the Raychaudhuri equation to investigate all of the effective energy conditions and then have tried to develope a general yet feasible $f(R,T)$ gravity model as an analytic solution, although keeping in mind that the Raychaudhuri equation has already been derived in the context of $f(R,T)$ gravity in a different way~\cite{Baffou}. The main purpose of adopting the Raychaudhuri equation in this scenario is to start as much as possible from an overall scenario because it is independent of the Einstein field equation. On the other hand, this theory can work equally well in the special form $f(R,T)=f_1(R)+f_2(T)$ because the Raychaudhuri equation only assumes the geometrical perspectives on the background of the Reimannian manifold. In this work, we have obtained a combination of powers of the Ricci scalar $R$ and trace of the energy-momentum tensor $T$ and finally integrate the Raychaudhuri equation for the choice of $f(R,T)=f_1(R)+f_2(T)$. The analytic solutions of two different models have been discussed and we have also checked all the energy conditions as well as stability analysis from those particular analytical solutions which are purely geometrical.

The paper is organized as follows. Sec. II deals with the relevant theory in $f(R,T)$ gravity. In Sec. III, we have briefly discussed the Raychaudhuri equation and presented the relevant energy conditions. The fourth Sect. addressed the energy conditions as well as an overview of the Raychaudhuri equation in $f(R,T)$ gravity.  The general construction of $f(R,T)$ models for two different examples such as the matter-dominated and the dark energy-dominated universe has been done in Sec. V, assuming the power law expansion. Section VI are kept for some verification schemes, such as VI-A deals with the viability analysis of this theory whereas subsection VI-B deals with the analysis of energy conditions. Finally, we put an end at Sec. VII with concluding remarks.

\section{$f(R,T)$ gravity theory}
The action for $f(R,T)$ gravity has the form~\cite{Harko}:
\ben
\A=\int d^4 x\s{-g}\Big[\frac{1}{2\kappa}f(R,T)+\L\Big],
\label{1}
\een 
where $f(R,T)$ is an analytic function of Ricci scalar $R$ and trace of the energy-momentum tensor $T$, $\L$ is the matter Lagrangian density. Here $\k=8\pi G$, $G$ being the universal gravitational constant. 

The energy-momentum tensor $T_{\a\b}$ is:
\ben
T_{\a\b}=-\frac{2}{\s{-g}}\frac{\d(\s{-g}\L)}{\d g^{\a\b}},
\label{2}
\een
and the trace of it is $T=T_{\a\b}~g^{\a\b}$, where $g^{\a\b}$ is the usual gravitational metric.

Assuming the matter dependency of Lagrangian density only on the metric tensor $g_{\a\b}$ and not on its derivative, we can write
\ben
T_{\a\b}=g_{\a\b}\L -2\frac{\d\L}{\d g^{\a\b}}.
\label{3}
\een

Using the method of metric formalism, we get the field equation by varying the corresponding action of Eq. (\ref{1}) as
\ben
&&f_{R}R_{\a\b}-\frac{1}{2}g_{\a\b}f(R,T)+\Big(g_{\a\b}\sq -\nb_\a \nb_\b\Big)f_{R}\non\\
&&=\kappa T_{\a\b}-f_{T}\Big(T_{\a\b}+\T_{\a\b}\Big),
\label{4}
\een
where $f_{R}$ and $f_{T}$ denote the derivative of $f(R,T)$ with respect to $R$ and $T$ respectively. Here $\sq=\nb^{\mu}\nb_{\mu}$, known as the D'Alembertian operator and $\nb_\a$ is the covariant derivative associated with the metric tensor $g_{\a\b}$. Also, here $\T_{\a\b}$ is defined as~\cite{Harko}
\ben
&& \T_{\a\b}\equiv g^{\mu\nu}\frac{\d T_{\mu\nu}}{\d g^{\a\b}}\non\\
&&=g_{\a\b}\L -2T_{\a\b}-2g^{\mu\nu}\frac{\d^2 \L}{\d g^{\a\b}\d g^{\mu\nu}}.
\label{5}
\een

Considering the matter Lagrangian $\L=+p$~\cite{Harko}, we have from Eq. (\ref{5}), the following expression
\ben
\T_{\a\b}=-2T_{\a\b}+p g_{\a\b}.
\label{6}
\een

The above field equation (\ref{4}) can be expressed in $f(R,T)$ gravity in the form of Einstein tensor in addition to the {\it effective energy-momentum} tensor $\Tilde{T}_{\a\b}$ which withholds the curvature contribution as~\cite{Harko},
\ben
G_{\a\b}\equiv R_{\a\b}-\frac{1}{2}g_{\a\b}R=\Tilde{T}_{\a\b},
\label{7}
\een
where
\ben
&& \Tilde{T}_{\a\b}=\frac{1}{f_R}\Big[\big(\kappa +f_{T}\big)T_{\a\b}+\frac{1}{2}g_{\a\b}\big(f(R,T)-Rf_{R}\big)\non\\
&&-f_{T} p g_{\a\b}+D_{\a\b}f_{R}\Big]\label{8}.
\een

Here the operator $D_{\a\b}$ is given by $D_{\a\b}\equiv(\nb_\a\nb_\b-g_{\a\b}\sq)$. Eq. (\ref{7}) is the modified Einstein equation in $f(R,T)$ gravity. The extra terms arise in the effective energy-momentum tensor due to the presence of $f_{R}$ and $f_{T}$, which are not zero here. So, the effective gravitational coupling, i.e., the terms due to the coupling of $R$ and $T$ with geometry has non-zero effect rather than the usual gravitational theory.

We have also the following expression of the Ricci tensor as
\ben
&& R_{\a\b}=\frac{1}{f_{R}}\Big[\big(\kappa +f_{T}\big)T_{\a\b}-f_{T}pg_{\a\b}+\frac{1}{2}g_{\a\b}f(R,T)\non\\
&&+D_{\a\b}f_{R}\Big].
\label{9}
\een

\section{Raychaudhuri equation and Energy conditions}
The Raychaudhuri equation for timelike geodesic congruence with tangent vector filed $v^\a$ and affine parameter $\tau$ is given by~\cite{Ray55,Ray57,Ray,Kar,Poisson}
\ben
\frac{d\th}{d\tau}+(\nb_\a v^\b)(\nb_\b v^\a)=-R_{\g\b}v^\g v^\b,
\label{10}
\een
and the quantity $(\nb_\a v^\b)(\nb_\b v^\a)$ can be written as
\ben
(\nb_\a v^\b)(\nb_\b v^\a)=2{\sg}^2-2\o^2+\frac{1}{3}{\th}^2
\label{11}
\een
where for affinely parameterised geodesic equation we set the acceleration term $v^\a\nb_\a v^\b=0$~\cite{Ray,Poisson}. Here, $\th=\nb_\a v^\a$ is called the scalar expansion or volume expansion,   $\sg_{\a\b}=\frac{1}{2}\Big(\nb_\b v_\a+\nb_\a v_\b\Big)-\frac{1}{3}{\th} h_{\a\b}$ is called the symmetric shear whereas anti-symmetric rotation is given as $\o_{\a\b}=\frac{1}{2}\Big(\nb_\b v_\a-\nb_\a v_\b\Big)$. Also, $\nb_\b v_\a=(\sg_{\a\b}+\o_{\a\b}+\frac{1}{3}{\th} h_{\a\b})$, $2{\sg}^2=\sg_{\a\b}\sg^{\a\b}$, $2\o^2=\o_{\a\b}\o^{\b\a}$ with the three-dimensional hypersurface metric $h_{\a\b}=g_{\a\b}-v_\a v_\b$.

So, finally the Raychaudhuri equation for the congruence of timelike geodesics, i.e., Eq. (\ref{10}) takes the form as
\ben
\frac{d\th}{ds}=-\frac{1}{3}{\th}^2 -2{\sg}^2+2\o^2 -R_{\a\b}v^\a v^\b.
\label{12}
\een

On the other hand, the Raychaudhuri equation for the congruence of null geodesics with a vector field $k^{\mu}$ is defined as
\ben
\frac{d\th}{ds}=-\frac{1}{2}{\th}^2 -2{\sg}^2+2\o^2 -R_{\a\b}k^\a k^\b.
\label{13}
\een

It is remarkable to say that the Raychaudhuri equation is completely geometrical in nature and it is independent of any other gravitational theories. By making a connection between the Raychaudhuri equation and the Einstein field equations, one can obtain the physical conditions on the energy-momentum tensor.

It is noteworthy to mention that Raychaudhuri's work~\cite{Ray55,Ray57} is predominantly focused on the field of cosmology. He assumes that the Universe can be represented by a time-dependent geometry but does not assume homogeneity or isotropy at the outset. In fact, one of his aims is to see whether non-zero rotation (spin), anisotropy (shear) and/or a cosmological constant can succeed in avoiding the initial singularity. For the purpose of studying the large-scale structure of the cosmos, we assume throughout this work that the Universe is homogeneous as well as isotropic. For choices of the Universe's homogeneity and isotropy, we can assume that the shear, which is a spatial tensor, is zero, i.e., ${\sg}^2\equiv \sg_{\a\b}\sg^{\a\b}=0$ and from the Frobenius' theorem~\cite{Poisson}, the rotation tensor $\o^2\equiv \o_{\a\b}\o^{\a\b}=0$. So, the constraints for gravity to be attractive in nature must satisfy the following conditions:\\ 
\ben
R_{\a\b}v^\a v^\b\geq 0~(SEC),\label{14}\\
R_{\a\b}k^\a k^\b\geq 0~(NEC).\label{15}
\een

By using the usual Einstein field equations and Eq. (\ref{14}), the statement of the strong energy condition with a timelike vector field $v^\a$ is given by
\ben
\big(T_{\a\b}-\frac{1}{2}T g_{\a\b}\big)v^\a v^\b\geq0.
\label{16}
\een

On the other hand, the null energy condition states that the energy density of any matter distribution, as measured by any observer with null vector field $k^\a$ in spacetime, must be non-negative. So from Eq. (\ref{15}), one can get
\ben
T_{\a\b}k^\a k^\b\geq0.
\label{17}
\een

Again, the weak energy condition makes the same statement as the null form, except that the lightlike vector field $k^\a$ is replaced by a timelike vector field $v^\a$. So one can achieve
\ben
T_{\a\b}v^\a v^\b\geq0.
\label{18}
\een

And finally, the dominant energy condition states that no signal can propagate faster than the speed of light. So matter should ﬂow along timelike or null world lines. That means the matter’s momentum density $-T^\a_\b v^\b$ as measured by an observer with four-velocity $v^\a$ must be timelike or null, vector ﬁeld.

If the cosmic perfect fluid-like energy-momentum tensor $T_{\a\b}$ with energy density $\r$, pressure $p$ and the four-velocity $v_\a$ is described as
\ben
T_{\a\b}=(\r+p)v_\a v_\b+p g_{\a\b},
\label{19}
\een 
and then the most familiar form of all energy conditions in GR is given as
\ben
&&\r+3p\geq 0~ {\text{and}}~\r+p\geq 0~(SEC),\label{20}\\
&&\r+p\geq 0~(NEC),\label{21}\\
&&\r\geq 0~\text{with}~\r+p\geq 0~(WEC),\label{22}\\
&&\r\geq 0~\text{with}~\r\pm p\geq 0~(DEC).\label{23}
\een

It should be mentioned that the violation of NEC leads to violation of other energy conditions. One can express the energy conditions in the modified theory of gravity, similar to GR by replacing the ordinary energy density $\r$ and pressure $p$ by the effective one, i.e., $\rf$ and $\pf$.

\section{Energy Conditions and Raychaudhuri equation in $f(R,T)$ gravity}

\subsection{Energy Conditions in $f(R,T)$ gravity}
Using Eqs. (\ref{7}), (\ref{8}) and (\ref{19}), we have the expression for trace of the effective energy-momentum tensor ($\kappa=1$), which is:
\ben
\Tilde{T}=\frac{1}{f_R}\Big[(1+f_T)(3p-\r)+2f(R,T)-2Rf_R\non\\
-4pf_T-3\sq f_R\Big].
\label{24}
\een

As we are dealing with $f(R,T)$ gravity, due to the interaction between matter and curvature which imposes an extra acceleration, the energy-momentum tensor of matter is not conserved~\cite{Sharif1,Sharif2}. That means the massive test particles does not follow a geodesic trajectory due to the presence of extra force.

The attractive nature of gravity must adhere to the following supplementary restriction within the framework of $f(R,T)$ gravity~\cite{Sharif2}
\ben
\dfrac{1+f_T}{f_R}>0.
\label{25}
\een

The inequality holds independently of the conditions derived from the Raychaudhuri equation, as stated in Eqs. (\ref{14}) and (\ref{15}).

Now we investigate all the energy conditions in the $f(R,T)$ gravity. In this study for simplicity, we consider flat Friedmann-Lemaitre-Robertson-Walker (FLRW) metric corresponding to a homogeneous and isotropic universe having the following line element
\ben
ds^2=-dt^2+a^2(t)d{\bf x}^2,
\label{26}
\een  
where $d{\bf x}^2=dr^2+r^2 d\O^2$ is the spatial part of the metric and $d\O^2=d\t^2 +\sin^2\t d\phi^2$ is the line element of unit two sphere. Here $a(t)$ is the cosmological scale factor. So, assuming Eqs. (\ref{19}) and (\ref{26}) and also using Eqs. (\ref{8}) and (\ref{24}) the corresponding effective energy density $\rf$ and pressure $\pf$ can be obtained in this theory as
\ben
\rf=\frac{1}{f_R}\Big[\r+(\r+p)f_T-\frac{1}{2}\big(f-Rf_R\big)-3Hf_{RR}\dot{R}\Big],
\nonumber\\
\label{27}
\een

\ben
\pf=\frac{1}{f_R}\Big[p+\frac{1}{2}\big(f-Rf_R\big)+2Hf_{RR}\dot{R}+f_{RR}\ddot{R}\nonumber\\+f_{RRR}\dot{R}^2\Big],
\label{28}
\een
where  $H=\frac{\dot{a}}{a}$ is the Hubble parameter, $f_{RR}$ and $f_{RRR}$ being the second and third order derivative of $f(R,T)$ with respect to $R$ and $\dot{R}$ and $\ddot{R}$ being the first and second order time derivative of Ricci scalar $R$, respectively.  Note that in the above equations under the consideration of $f(R,T)=R$ one can get $\Tilde{\rho}=\rho$ and $\Tilde{p}=p$, which are the usual case of Einstein's gravity. 

The strong energy condition in this modified theory can be achieved by using Eqs. (\ref{7}) and (\ref{14}) as
\ben
\Tilde{T}_{\a\b}v^\a v^\b-\frac{1}{2}\Tilde{T}\geq 0.
\label{29}
\een

So, from Eqs. (\ref{27}), (\ref{28}) and (\ref{29}), it follows the SEC in $f(R,T)$ gravity which is
\ben
\rf+3\pf=\frac{1}{f_R}\Big[\r+3p+(\r+p)f_T+f-Rf_R\non\\
+3\big(Hf_{RR}\dot{R}+f_{RR}\ddot{R}+f_{RRR}\dot{R}^2\big)\Big]\geq 0.
\label{30}
\een

The {\bf NEC} in $f(R,T)$ gravity in the form $\Tilde{T}_{\a\b}k^\a k^\b\geq 0$ results the following inequality
\ben
\rf+\pf=\frac{1}{f_R}\Big[\r+p+(\r+p)f_T-Hf_{RR}\dot{R}\non\\
+f_{RR}\ddot{R}+f_{RRR}\dot{R}^2\Big]\geq 0.
\label{31}
\een

It should be noted that the standard form of the NEC and SEC can be found in GR as a special case by taking $f(R,T)=R$. Also, we have mentioned that the origin of the derivation of SEC and NEC given by Eqs. (\ref{30}) and (\ref{31}) are basically the Raychaudhuri equation. In determining the weak energy condition and the dominant energy condition, we can consider the modified form of energy conditions in standard GR which are obtained under the transformations of $\r \rightarrow \rf$ and $p\rightarrow \pf$.

So, the WEC requires the condition $\rf \geq 0$ with the condition of Eq. (\ref{31}) which follows the given inequality
\ben
\rf=\frac{1}{f_R}\Big[\r+(\r+p)f_T-\frac{1}{2}\big(f-Rf_R\big)-3Hf_{RR}\dot{R}\Big]\geq 0.\non\\
\label{32}
\een

Finally the DEC is agreed by meeting the inequalities Eqs. (\ref{31}) and (\ref{32}) and the condition
\ben
\rf-\pf=\frac{1}{f_R}\Big[\r-p+(\r+p)f_T-f+Rf_R-5Hf_{RR}\dot{R}\non\\
-f_{RR}\ddot{R}-f_{RRR}\dot{R}^2\Big]\geq 0.~~\label{33}
\een

So, these are all the energy conditions in $f(R,T)$ gravity. Also, by neglecting the dependence on the trace of the energy-momentum tensor in the action formalism, we can have the energy conditions in $f(R)$ gravity~\cite{Santos1,Santos2}.

\subsection{Raychaudhuri equation in $f(R,T)$ gravity}
The scalar expansion or volume expansion $\th$ in Eq. (\ref{11}) can be defined as the fractional rate of change in the cross sectional volume of the congruence, denoted by $\th=\frac{1}{\d V}\frac{d}{d\tau}(\d V)$~\cite{Poisson,Ray}. On the other hand from cosmological background $\th$ can be measured in a comoving Lorentz frame by the following definition~\cite{Ray,Kar,Sebas,Poisson} as
\ben
&&\th=\nb_\a v^\a\equiv\frac{1}{\sqrt{-g}}\p_\a(\sqrt{-g}v^\a)=\frac{1}{a^3}\p_0(a^3)\non\\
&&=\frac{3}{a}\frac{da}{dt}\label{34}.
\een

The gravity effect $R_{\a\b} v^\a v^\b$ plays a crucial role in the construction of the Raychaudhuri equation based on Riemannian geometry. So we have the following equation after contracting Eq. (\ref{9}) with $v^\a v^\b$ for such a metric, i.e., Eq. (\ref{26})
\ben
R_{\a\b}v^\a v^\b &&=\frac{1}{f_{R}}\Big[\Big(1+f_{T}\Big)\r+f_{T}p-\frac{1}{2}f(R,T)\non\\
&&-3Hf_{RR}\dot{R}\Big],
\label{35}
\een
where $v^\a v^\b D_{\a\b}f_{R}=-3Hf_{RR}\dot{R}$ has been achieved by taking the velocity tangent vector in a comoving Lorentz frame just for cosmological simplicity.
  
Finally, using Eqs. (\ref{34}) and (\ref{35}) in Eq. (\ref{11}) we get the Raychaudhuri equation in $f(R,T)$ gravity as harmonic oscillator equation with varying frequency~\cite{Kar,Tipler1,Tipler2}
\ben
&&\frac{d^2 a}{dt^2}+\frac{1}{3}\Big(R_{\a\b} v^{\a} v^{\b}+2\sg^2-2\o^2\Big)a=0
\label{36}.
\een

In some references~\cite{Lawrence}, $\th$ may be also identified as the derivative of the geometric entropy $S$, i.e., $S=\ln a$ and $a$ can be termed as an effective deviation of congruence of the geodesic. But in our case, $a(t)$ simply gives us the history of the cosmic evolution of the universe and from that cosmological perspective, $\frac{d\th}{d\tau}$ is simply the acceleration term relating to the standard cosmology.

For the zero shear and zero vorticity (rotation) under the consideration of maximally symmetric spacetime metric (i.e., the spacetime is both homogeneous and isotropic)~\cite{Weinberg}, \textcolor{red}{using Eqs. (\ref{12}), (\ref{34}) and (\ref{35}),} we can have a particular case of generalized Raychaudhuri equation as~\cite{Baffou}
\ben
\frac{\ddot{a}}{a}=\frac{1}{f_{R}}\Big[-\frac{\r}{3}-\frac{f_{T}}{3}(\r+p)+\frac{1}{6}f(R,T)+Hf_{RR}\dot{R}\Big].\non\\
\label{37}
\een

We get the above equation with a pressure term as we are dealing with $f(R,T)$ gravity and what is obvious.

This equation is an alternative version of the Raychaudhuri equation (\ref{12}) within the framework of $f(R,T)$ gravity theory, which is derived from Eq. (\ref{36}).
At this juncture, we would like to discuss the well-known Friedmann equation. The $(00)$ and $(ii)$ components of the Einstein equation provide two equations known as the Friedmann equations~\cite{Friedmann}, which are named after the Russian mathematician and meteorologist (1922). In both of these equations, Friedmann assumes homogeneity and isotropy of the cosmos where the $(00)$ component gives the first Friedmann equation whereas the $(ii)$ component gives the second Friedmann equation. In $f(R,T)$ gravity, two Friedmann equations can be expressed as (for $k=0$) \cite{Baffou,Baffou1}:
\ben
H^2=\frac{1}{f_R}\Big[\rho+f_{T}(\rho+p)-3Hf_{RR}\dot{R}+\frac{1}{2}(Rf_{R}-f)\Big],
\label{eq1}
\een
\ben
-2\dot{H}-3H^2=&&\frac{1}{f_{R}}\Big[ p+\frac{1}{2}(f-Rf_{R})+2Hf_{RR}\dot{R}\nonumber\\
&&+f_{RRR}\dot{R}^2+f_{RR}\ddot{R}\Big].
\label{eq2}
\een
If we consider $H^2=\tilde{\rho}$ and $-2\dot{H}-3H^2=\tilde{p}$ then one can get Eq. (\ref{27}) and Eq. (\ref{28}) for ($\kappa^2=1$). The first Friedmann equation is the primary equation that governs the expansion of the Universe \cite{Weinberg1,Ta-Pei}. The energy conservation equation, derived from the combination of the first and second Friedmann equations, is expressed as:  
\ben
\dot{\tilde{\rho}}+3H (\tilde{\rho}+\tilde{p})=0.
\label{eq3}
\een
By utilizing the standard forms of the modified Friedmann equations (\ref{eq1}) and (\ref{eq2}), along with the modified energy conservation equation (\ref{eq3}) under the $f(R,T)$ gravity model for a flat Universe~\cite{Baffou,Baffou1}, it can be readily verified that the aforementioned generalized Raychaudhuri equation (\ref{37}) can easily be derived.

\section{A construction of $f(R,T)$ gravity model}
We are now trying to build a broad representation of $f(R,T)$ gravity based on the assumption that the cosmological scale factor varies with time in a power law form. This is done so that the model can mimic the behavior of the $\Lambda$-CDM model in GR. In this formalism, we take the form of a scale factor as
\ben
a(t)=Ct^m~; m>0,
\label{38}
\een
where $C$ is a constant. 

For the metric mentioned in Eq. (\ref{26}) and for the choice of the scale factor mentioned in Eq. (\ref{38}), one can express the Ricci scalar and the time derivative of the Ricci scalar as a function of scale factor, i.e.,
\ben
R=6\frac{\dot{a}^2}{a^2}+6\frac{\ddot{a}}{a}=6\Big(\frac{C}{a}\Big)^{\frac{2}{m}}\Big(2m^2-m\Big),
\label{39}
\een
and
\ben
\dot{R}=-2R\Big(\frac{C}{a}\Big)^{\frac{1}{m}}.
\label{40}
\een

To demonstrate the model's general solution, we describe all of the physical quantities in Eq. (\ref{37}) in terms of the scalar curvature $R$, which allows us to solve the differential equation of $f(R,T)$.

\subsection{A Solution of $f(R,T)=f_1(R)+f_2(T)$ Gravity Model}

\subsubsection{Matter Dominated Era}
If we assume the conservation of the energy-momentum tensor corresponding to the fluid distribution given in Eq. (\ref{19}) independently, then the continuity equation is satisfied. Now we can take the matter-dominated era and the fluid pressure during this era is $p=0$ with $\r=\r_0 a^{-3}$ as the reference. So, with the help of Eqs. (\ref{38}), (\ref{39}) and (\ref{40}), we can write Eq. (\ref{37}) in terms of $R$ as
\ben
&& R^2 f_{RR}+\frac{1}{2}(m-1)R f_{R}-\frac{1}{2}(2m-1)f(R,T)\non\\
&&=-\frac{\r_0}{a^3}(2m-1)\Big(1+f_{T}\Big).
\label{41}
\een

For matter-dominated era and $T=-\r$ (from the energy conservation we know that $T=-\r+3p$), one can express the above equation for the choice of $f(R,T)=f_1(R)+f_2(T)$ in the following way:
\ben
&& R^2 f_1''(R)+\frac{(m-1)}{2}R f_1'(R)-\frac{(2m-1)}{2}f_1(R)+A R^{\frac{3m}{2}}\non\\
&&-\frac{(2m-1)}{2}f_2(T)-\Big(2m-1\Big)T f_2'(T)=0,
\label{42}
\een
where $f_1'(R),~f_1''(R)$ denote the first and second order derivative of $f(R,T)$ w.r.t. $R$ respectively and $f_2'(T)$ denotes the first-order derivative of $f(R,T)$ w.r.t. the trace of the energy-momentum tensor $T$ and constant $A$, which is dependent on parameter $m$, is given by
\ben
A=\frac{\r_0}{C^3}\Big(2m-1\Big)\Big(\frac{1}{12m^2-6m}\Big)^{\frac{3m}{2}}\label{43}.
\een

It should be mentioned that in Eq. (\ref{41}), there is a time dependence through the scale factor $a(t)$, but in Eq. (\ref{42}), we have eliminated the direct dependence of the scale factor by using Eq. (\ref{39}), where $a(t)$ can be written as in terms of Ricci scalar $R$. So that the modified Raychaudhuri Eq. (\ref{37}) can be written in terms of $R$ and $T$ and their respective derivatives in Eq. (\ref{42}), i.e., this equation is free from direct dependence of time.

As $f_1(R)$ and $f_2(T)$ are two explicitly independent functions of $R$ and $T$ in Eq. (\ref{42}), we can use the separation of the variable method to get  the general analytic solution of $f(R,T)$ from Eq. (\ref{42}) which can be written as
\ben
f(R,T)=C_1 R^\a + C_2 R^\b + B R^\d +\frac{C_3}{\s T},
\label{44}
\een
where $C_1,~C_2$ and $C_3$ are arbitrary integration constants. The other constants $B,~\a,~\b,~\d$ are as follows:
\ben
&& B=-\frac{4 A}{(12m^2-13m+2)},\label{45}\\
&&\a=\frac{1}{4}\Big[3-m+\s{(1+10m+m^2)}\Big],\label{46}\\
&&\b=\frac{1}{4}\Big[3-m-\s{(1+10m+m^2)}\Big],\label{47}\\
&&\d=\frac{3m}{2}.\label{48}
\een 

It is to be noted that for the real values of $A$ and $B$, we are bound to vary $m$ from $m=0.5$ to get a real cosmological model in $f(R,T)$ gravity theory. 

There are three different powers of $R$ in the solution of $f(R,T)$ namely, $\a,~\b$ and $\d$. Figures 1,  2 and 3, respectively, show the variations of these three parameters with $m$. Note that in Figs. 1, 2 and 3, we have taken the range of $m$ from $0.5$ for the reason discussed above. However, we can observe in Fig. 1 that $\a$ increases as $m$ increases. It will become constant at almost 2 for a larger $m$ value, and the first term of Eq. (\ref{44}) will behave as $R^2$. Figure 2 depicts $\b$, which begins at $0.0$ and then becomes negative, as $m$ increases, $\b$ becomes more negative. As expected for $\d$, it varies linearly with $m$ and goes on increasing in Fig. 3. 

Note from Fig. 2 that as $m$ increases from $0.5$, $\b$ becomes more negative, implying that the second term of Eq. (\ref{44}) is more negative curvature, and at the point $R=0$, it is undefined. In order to avoid the discontinuity of the $R^{\b}$ term in Eq. (\ref{44}), we can choose the arbitrary constant $C_{2}=0$ in the region where $m>0.5$. Therefore, after $m=0.5$, the only contribution to the curvature comes from $\a~\&~\d$, not from $\b$.

\begin{figure}
\centering
\includegraphics[width=6.0cm, height=6.0cm]{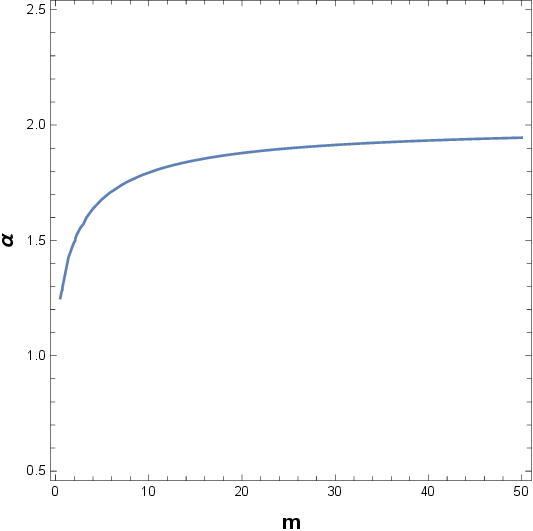}
 \caption{Variation of $\a$ with $m$.}\label{Fig1}
\end{figure}

\begin{figure}
\centering
\includegraphics[width=6.0cm, height=6.0cm]{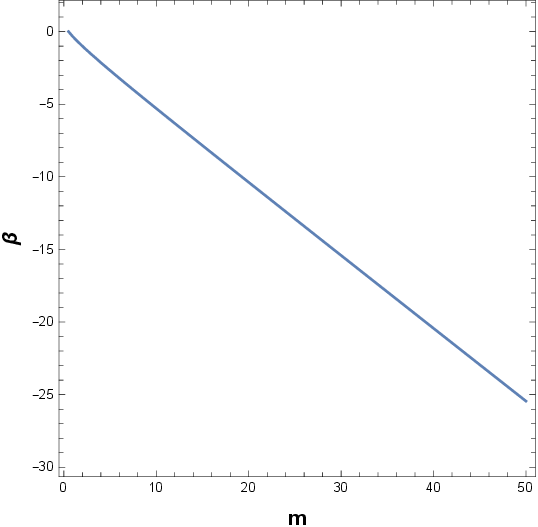}
 \caption{Variation of $\b$ with $m$.}\label{Fig2}
\end{figure}

\begin{figure}
\centering
\includegraphics[width=6.0cm, height=6.0cm]{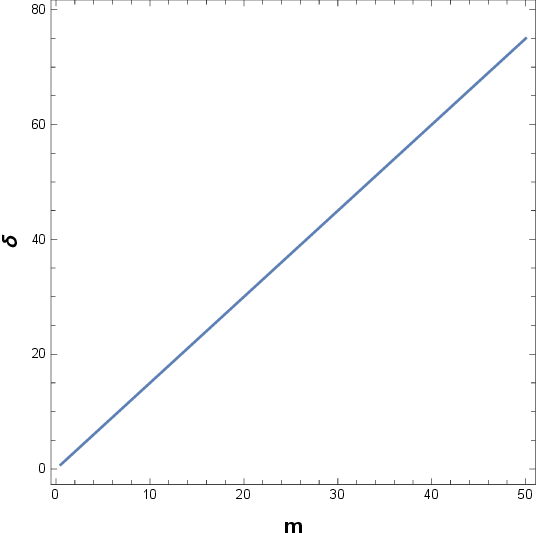}
 \caption{Variation of $\d$ with $m$.}\label{Fig3}
\end{figure}

\subsubsection{Accelerating Dark Energy-Dominated Era}
We know for dark energy-dominated universe, $w\simeq -1$ \cite{Weinberg1} as $p=w\rho$. But for an accelerated universe, one should have $\ddot{a}>0$. So one can choose $w$ as $-1<w\leq -\frac{1}{3}$ for dark energy in which the universe expands as a power law \cite{Jose}. We consider the energy density varies with scale factor as~\cite{Jose} $\rho=\rho_{0} a^{-3(1+w)}=\rho_{0} a^{-\frac{3}{2}}$ for chosen $w=-0.5$. It should be noted that $w=-1$ indicates that $\rho$ is constant.

This is simply a fluid representation of the cosmological constant. $\rho$ is an increasing function of $a$ if $w\leq-1$. On the other hand, $\rho$ is a decreasing function of $a$ if $w\geq-1$. It is worth noting that the WMAP satellite team~\cite{Spergel} reported that $w\leq-0.78$, which is compatible with a cosmological constant.

As per the above case (matter-dominated era), for the conservation of energy-momentum tensor ($T=-\rho+3p$) corresponding to the fluid distribution given in Eq. (\ref{19}) and from Eq. (\ref{37}), with the help of Eqs. (\ref{38}), (\ref{39} and (\ref{40}), one can write
\ben
&& R^2 f_{RR}+\frac{1}{2}(m-1)R f_{R}-\frac{1}{2}(2m-1)f(R,T)\non\\
&&=-\r(2m-1)\Big(1+\frac{1}{2}f_{T}\Big),
\label{49}
\een
where we use $w=-0.5$ \cite{Jose} in the dark energy-dominated accelerating universe and then  $T=-2.5\r$. For the choice of $f(R,T)=f_1(R)+f_2(T)$, Eq. (\ref{49}) takes the form as
\ben
&& R^2 f_1''(R)+\frac{(m-1)}{2}R f_1'(R)-\frac{(2m-1)}{2}f_1(R)+{A}^{\prime} R^{\frac{3m}{4}}\non\\
&&-\frac{(2m-1)}{2}f_2(T)-\frac{T(2m-1)}{5}f_2'(T)=0,
\label{50}
\een
where the constant ${A}^{\prime}$ is given by
\ben
A^{\prime}=\frac{\r_0}{C^{\frac{3}{2}}}\Big(2m-1\Big)\Big(\frac{1}{12m^2-6m}\Big)^{\frac{3m}{4}}\label{51}.
\een

Therefore, the analytic solution of $f(R,T)$ gravity model for dark energy-dominated accelerated expansion can be written as
\ben
f(R,T)={C_1}^{\prime} R^{\a_d} + {C_2}^{\prime}  R^{\b_d} + {B}^{\prime}  R^{\d_d} +\frac{{C_3}^{\prime}}{T^{\frac{5}{2}}},
\label{52}
\een
where ${C_1}^{\prime} ,~{C_2}^{\prime} $ and ${C_3}^{\prime} $ are arbitrary integration constants and $~\a_d,~\b_d$ are same as $~\a,~\b$ as in the matter-dominated case. The other constants ${B}^{\prime},~\d_d$ are as follows:
\ben
&& B^{\prime} =-\frac{16 A^{\prime} }{(15m^2-34m+8)},\label{53}\\
&&\d_d=\frac{3m}{4}.\label{54}
\een 

It is noteworthy that the contribution of matter in the case of dark energy-dominated accelerating universe has been reduced $(T^{-\frac{5}{2}})$ than the case of the pressure-less matter distribution region of the universe $(T^{-\frac{1}{2}})$ and it is obvious because the matter contribution has minimal effect in the present accelerating era.

Here also we are bound to vary $m$ from $m=0.5$ as mentioned in the previous case. As $\a_d\equiv\a$ and $\b_d\equiv\b$, the variation of $\a_d,~\b_d$ are the same as in Figs. 1 and 2. The variation of $\d_d$ with $m$ is shown in Fig. 4 and the combined story of $\a_d,~\b_d,~\d_d$ with $m$ is shown in Fig. 5. It should be again mentioned that the choice of the arbitrary constant ${C_2}^{\prime}=0$ in the region ($m>0.5$) to get the net contribution to the curvature from the parameters $\a_d~\&~\d_d$. 

\begin{figure}
\centering
\includegraphics[width=6.0cm, height=6.0cm]{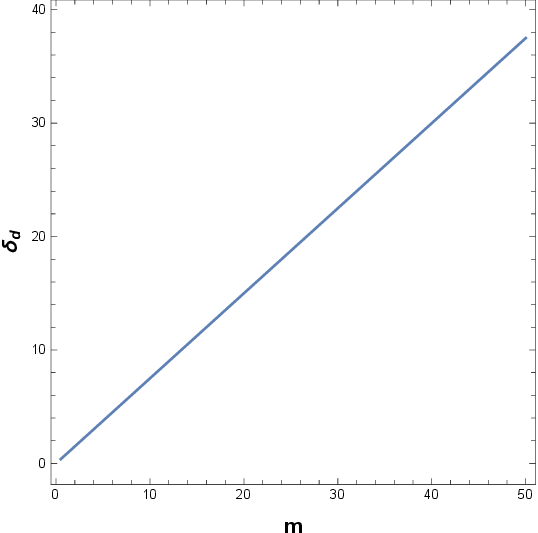}
 \caption{Variation of $\d_d$ with $m$.}\label{Fig5}
\end{figure}

\begin{figure}
\centering
\includegraphics[width=6.5cm, height=6.5cm]{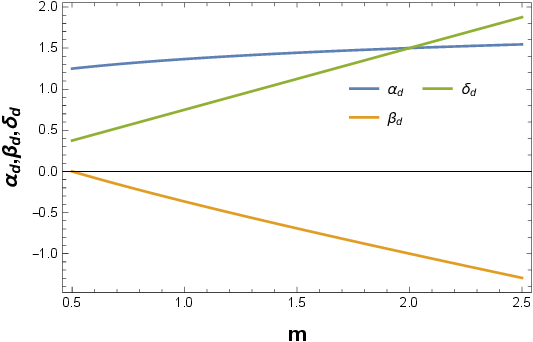}
 \caption{Variation of $\a_d,\b_d~\text{and}~\d_d$ with $m$.}\label{Fig6}
\end{figure}

\section{Analysis of the presented model}

\subsection{Viability of the physical features}
In our present work, we considered the scale parameter varies as a power law and constructed an analytical form of $f(R,T)$ gravity mentioned for the matter-dominated era (Eq. (\ref{44})) and for the accelerated dark energy-dominated era (Eq. (\ref{52})). But there remains a question, that how much viable our model is. To check this, we remind the works of several scientists~\cite{Saw,Silves,Sotirio,Sharif}, where they have mentioned that the viability conditions for any $f(R)$ or $f(R,T)$ theory are $f_{R}(R,T)> 0$, $f_{RR}(R,T)>0$, $R\geq R_{0}$ and also $f_{T}(R,T)>0$~\cite{Sharif3}, where $R_{0}$ is the present Ricci scalar. It can be noted that the first condition allows us to achieve a positive effective gravitational constant $\Big(G_{eff}\equiv\frac{1}{f_{R}(R,T)}(G+\frac{f_{T}(R,T)}{8\pi})\Big)$ of Eq. (\ref{7})~\cite{Sotirio,Sharif3}, whereas the other conditions ensure the stability of the model~\cite{Sharif,Sharif3}. Also, in~\cite{Sharif}, the authors have mentioned that the viability condition remains the same for $f(R,T)$ gravity as in the case of $f(R)$ gravity.  For both the matter-dominated and the accelerating dark energy-dominated eras, Fig. 2 demonstrates us that at $R\to 0$, the second term will diverge as $\b$ is negative in Eqs. (\ref{44}) and (\ref{52}). So, we have to eliminate the second term and this can be easily done by choosing $C_2$(or$~{C_2}^{\prime})=0$ and $C_1$(or$~{C_1}^{\prime})>0$ and here we've chosen $C_1$(or$~{C_1}^{\prime})$ to be an integer. The variation of $\d$ is a little bit different in these two eras. This variation is steeper in matter-dominated epoch than the dark energy-dominated epoch (Figs. 3 and 4), i.e., for $m=1$, $\d=1.5$ but $\d_d=0.75$. From Eqs. (\ref{44}) and (\ref{52}), we can write the form of $f_{R}(R,T)$ and $f_{RR}(R,T)$ for matter-dominated epoch and dark energy-dominated epoch as
\ben
f_{R}(R,T)_{matter}=\delta B R^{\delta -1}+\alpha  C_1 R^{\alpha -1}+\beta  C_2 R^{\beta -1},
\label{55}
\een

\ben
f_{R}(R,T)_{de}=\delta_{d} B'  R^{\delta_{d} -1}+\alpha_{d}  C_1' R^{\alpha_{d} -1}+\beta_{d}  C_2' R^{\beta_{d} -1},~~
\label{56}
\een

\ben
f_{RR}(R,T)_{matter}&=&B \delta(\delta -1)  R^{\delta -2}+\alpha(\alpha -1)   C_1 R^{\alpha -2}\nonumber\\
&+& \beta(\beta -1)   C_2 R^{\beta -2},
\label{57}
\een

\ben
f_{RR}(R,T)_{de}&=&B' \delta_{d}(\delta_{d} -1)   R^{\delta_{d} -2}+\alpha_{d} (\alpha_{d} -1)  C_1' R^{\alpha_{d} -2}\nonumber\\
&+& \beta_{d}(\beta_{d} -1)   C_2 R^{\beta_{d} -2}.
\label{58}
\een 

Also, the expression for $f_{T}(R,T)$ can be written from the same Eqs. (\ref{44}) and (\ref{52}) as
\ben
f_{T}(R,T)_{matter}=-\frac{C_3}{2 T^{3/2}},
\label{59}
\een
and
\ben
f_{T}(R,T)_{de}=-\frac{5 C_3^{'}}{2 T^{7/2}}.
\label{60}
\een

As our model does not bound the value of $m$ at a higher limit, we choose it to have three different values, e.g., $1,~5~\&~20$.  Figures 6(a) and 6(b), Figs. 7(a) and 7(b) and Figs. 8(a) and 8(b) represent the variation of $f_{R}(R,T)$ with $R$ for three different values of $m(1,~5~\&~20)$, respectively. Figures 6(a), 7(a) and 8(a) have been plotted using Eq. (\ref{55}) whereas  Figs. 6(b), 7(b) and 8(b) have been plotted using Eq. (\ref{56}). 

On the other hand, Figs. 9(a) and 9(b), Figs. 10(a) and 10(b) and Figs. 11(a) and 11(b) represent the variation of $f_{RR}(R,T)$ with $R$ for the same. Figures 9(a), 10(a) and 11(a) have been plotted using Eq. (\ref{57}). Figures 9(b), 10(b) and 11(b) have been plotted using Eq. (\ref{58}). Figure 12(a) (using Eq. (\ref{59})) and Fig. 12(b) (using Eq. (\ref{60})) represent the variation of $f_{T}$ with $T$. It should be noted that, as previously indicated, we consider $C_2$(or$~{C_2}^{\prime})=0$ in all cases.

Also, we would like to mention that we have chosen the value of $\frac{\rho_{0}}{C^3}$ to be $1$ (Eq. (\ref{43})) to plot the figures for the matter-dominated epoch and also chosen the value of $\frac{\rho_0}{C^{3/2}}$ to be $1$ (Eq. (\ref{51})) for the dark energy-dominated epoch.

\begin{figure*}
\begin{minipage}[b]{0.4\linewidth}
\centering
 \begin{subfigure}[b]{0.9\textwidth}
        \includegraphics[width=\textwidth]{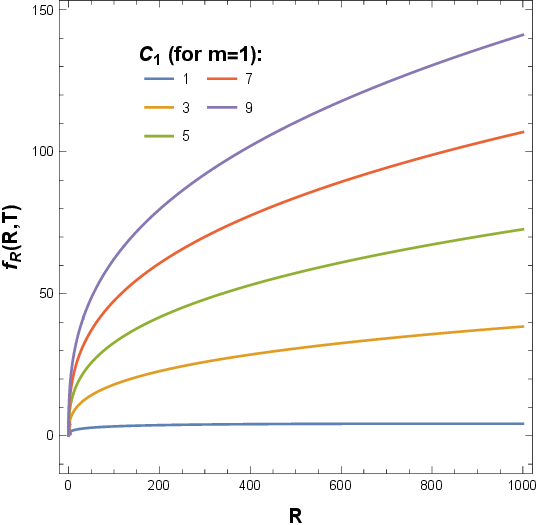}
        \caption{Matter dominated epoch}
        \label{6a}
    \end{subfigure}
\end{minipage}
\hspace{2cm}
\begin{minipage}[b]{0.4\linewidth}
\centering
 \begin{subfigure}[b]{0.9\textwidth}
        \includegraphics[width=\textwidth]{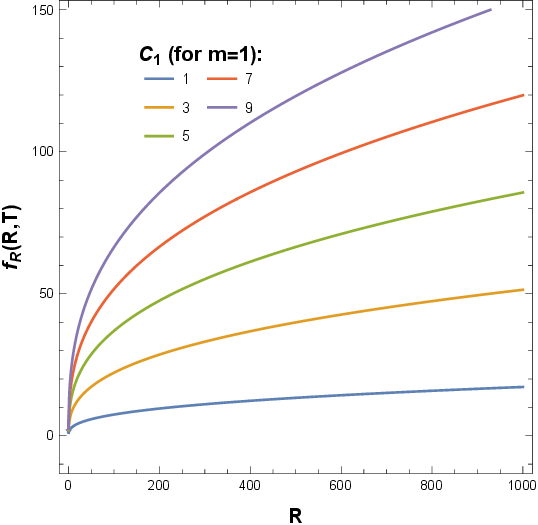}
        \caption{Dark energy dominated epoch}
        \label{6b}
    \end{subfigure}
\end{minipage}
\caption{Variation of $f_R(R,T)$ vs $R$ for $m=1$ in (a) matter and (b) dark energy dominated epoch}
\end{figure*}

\begin{figure*}
\begin{minipage}[b]{0.4\linewidth}
\centering
 \begin{subfigure}[b]{0.9\textwidth}
        \includegraphics[width=\textwidth]{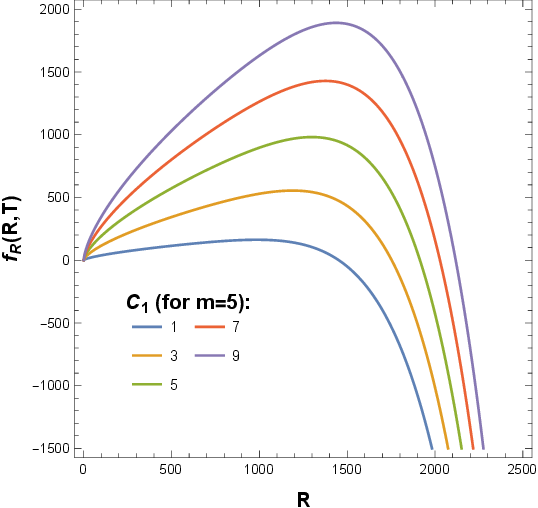}
        \caption{Matter dominated epoch}
        \label{7a}
    \end{subfigure}
\end{minipage}
\hspace{2cm}
\begin{minipage}[b]{0.4\linewidth}
\centering
 \begin{subfigure}[b]{0.9\textwidth}
        \includegraphics[width=\textwidth]{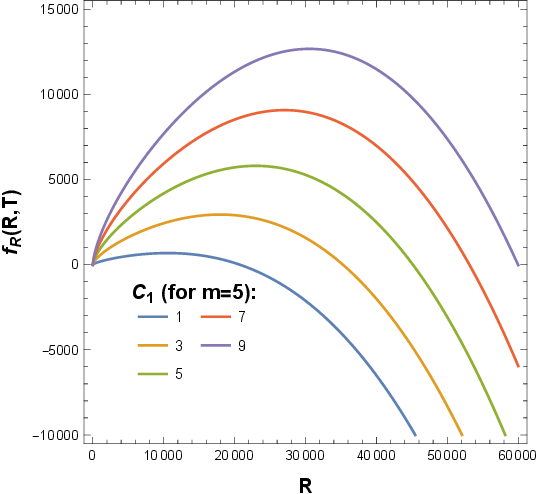}
        \caption{Dark energy dominated epoch}
        \label{7b}
    \end{subfigure}
\end{minipage}
\caption{Variation of $f_R(R,T)$ vs $R$ for $m=5$ in (a) matter and (b) dark energy dominated epoch}
\end{figure*}

\begin{figure*}
\begin{minipage}[b]{0.4\linewidth}
\centering
 \begin{subfigure}[b]{0.9\textwidth}
        \includegraphics[width=\textwidth]{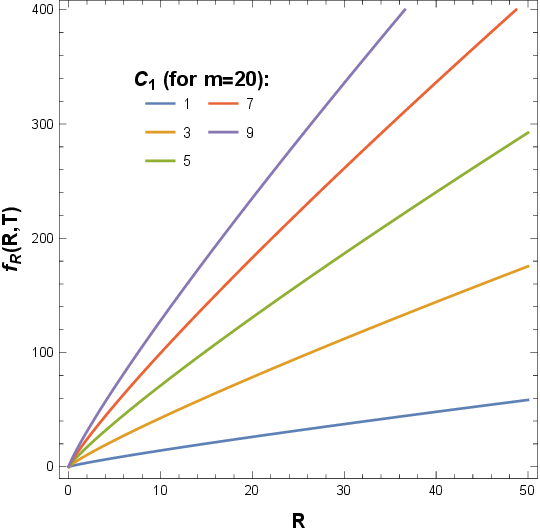}
        \caption{Matter dominated epoch}
        \label{8a}
    \end{subfigure}
\end{minipage}
\hspace{2cm}
\begin{minipage}[b]{0.4\linewidth}
\centering
 \begin{subfigure}[b]{0.9\textwidth}
        \includegraphics[width=\textwidth]{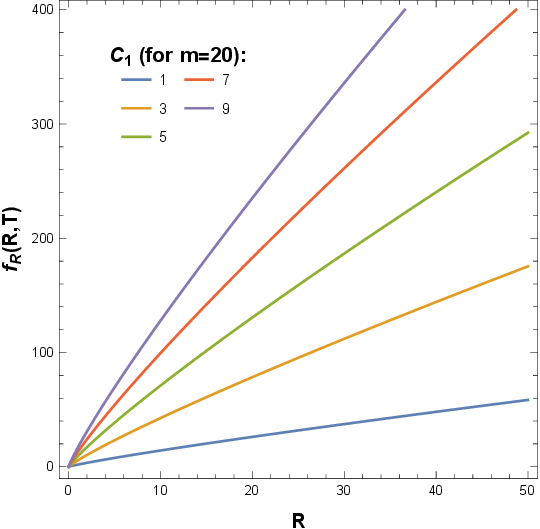}
        \caption{Dark energy dominated epoch}
        \label{8b}
    \end{subfigure}
\end{minipage}
\caption{Variation of $f_R(R,T)$ vs $R$ for $m=20$ in (a) matter and (b) dark energy dominated epoch}
\end{figure*}

\begin{figure*}
\begin{minipage}[b]{0.4\linewidth}
\centering
 \begin{subfigure}[b]{0.9\textwidth}
        \includegraphics[width=\textwidth]{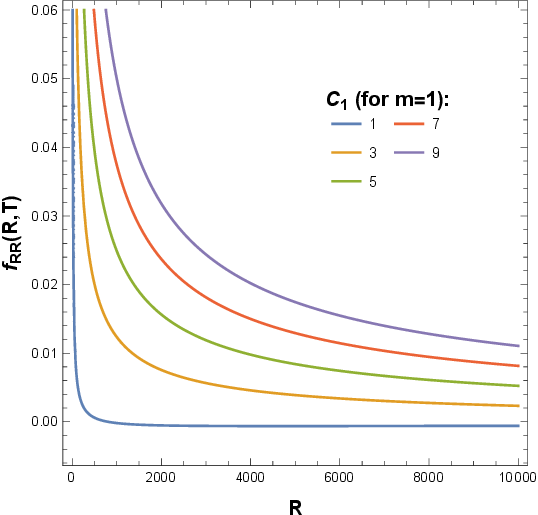}
        \caption{Matter dominated epoch}
        \label{9a}
    \end{subfigure}
\end{minipage}
\hspace{2cm}
\begin{minipage}[b]{0.4\linewidth}
\centering
 \begin{subfigure}[b]{0.9\textwidth}
        \includegraphics[width=\textwidth]{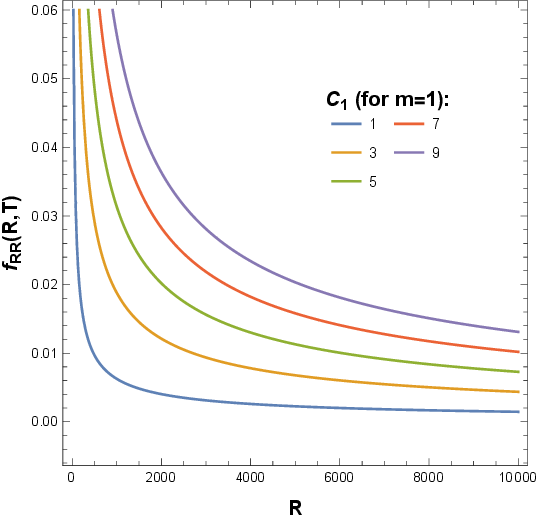}
        \caption{Dark energy dominated epoch}
        \label{9b}
    \end{subfigure}
\end{minipage}
\caption{Variation of $f_{RR}(R,T)$ vs $R$ for $m=1$ in (a) matter and (b) dark energy dominated epoch}
\end{figure*}

\begin{figure*}
\begin{minipage}[b]{0.4\linewidth}
\centering
 \begin{subfigure}[b]{0.9\textwidth}
        \includegraphics[width=\textwidth]{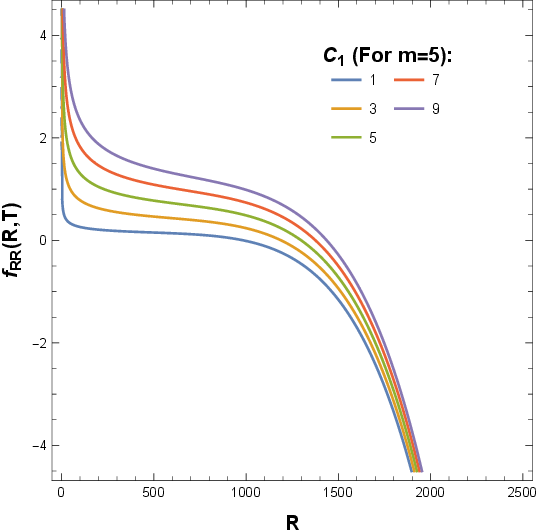}
        \caption{Matter dominated epoch}
        \label{10a}
    \end{subfigure}
\end{minipage}
\hspace{2cm}
\begin{minipage}[b]{0.4\linewidth}
\centering
 \begin{subfigure}[b]{0.9\textwidth}
        \includegraphics[width=\textwidth]{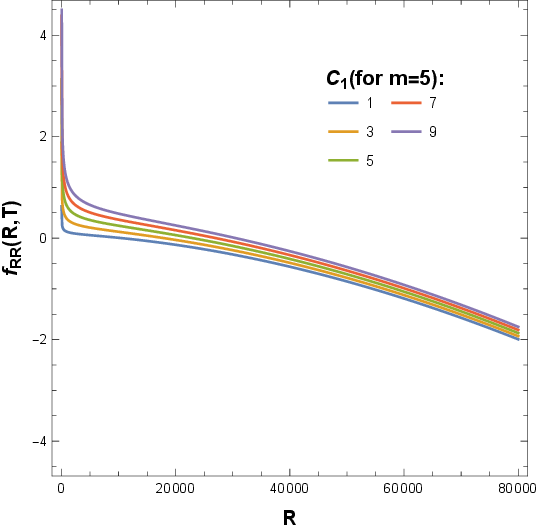}
        \caption{Dark energy dominated epoch}
        \label{10b}
    \end{subfigure}
\end{minipage}
\caption{Variation of $f_{RR}(R,T)$ vs $R$ for $m=5$ in (a) matter and (b) dark energy dominated epoch}
\end{figure*}

\begin{figure*}
\begin{minipage}[b]{0.4\linewidth}
\centering
 \begin{subfigure}[b]{0.9\textwidth}
        \includegraphics[width=\textwidth]{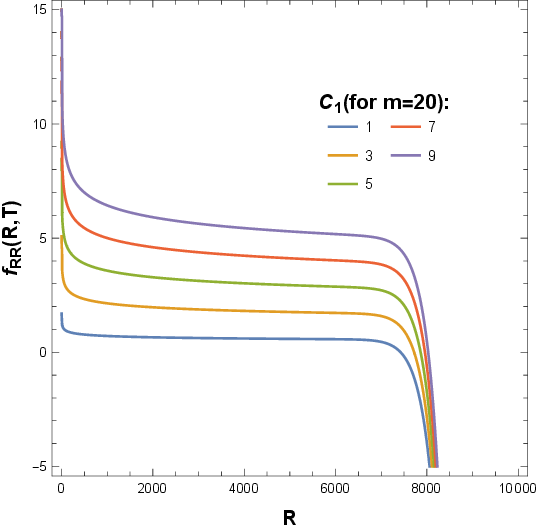}
        \caption{Matter dominated epoch}
        \label{11a}
    \end{subfigure}
\end{minipage}
\hspace{2cm}
\begin{minipage}[b]{0.4\linewidth}
\centering
 \begin{subfigure}[b]{0.9\textwidth}
        \includegraphics[width=\textwidth]{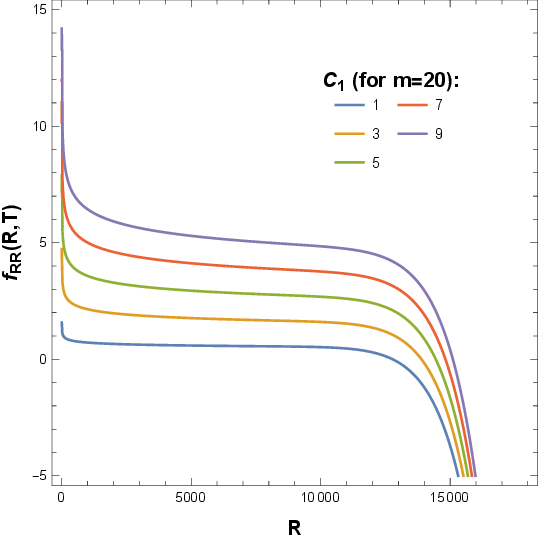}
        \caption{Dark energy dominated epoch}
        \label{11b}
    \end{subfigure}
\end{minipage}
\caption{Variation of $f_{RR}(R,T)$ vs $R$ for $m=20$ in (a) matter and (b) dark energy dominated epoch}
\end{figure*}

\begin{figure*}
\begin{minipage}[b]{0.4\linewidth}
\centering
 \begin{subfigure}[b]{0.9\textwidth}
        \includegraphics[width=\textwidth]{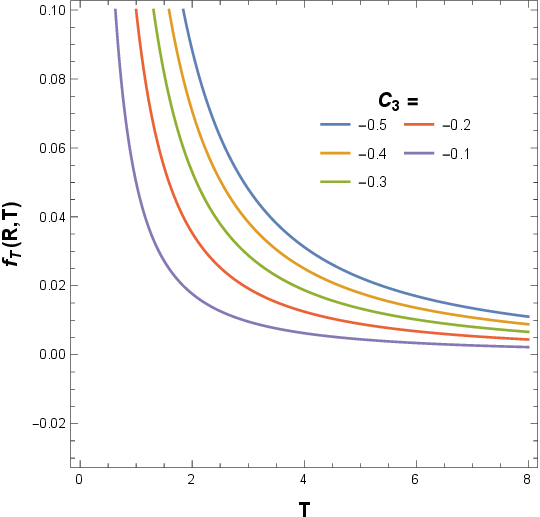}
        \caption{Matter dominated epoch}
        \label{12a}
    \end{subfigure}
\end{minipage}
\hspace{2cm}
\begin{minipage}[b]{0.4\linewidth}
\centering
 \begin{subfigure}[b]{0.9\textwidth}
        \includegraphics[width=\textwidth]{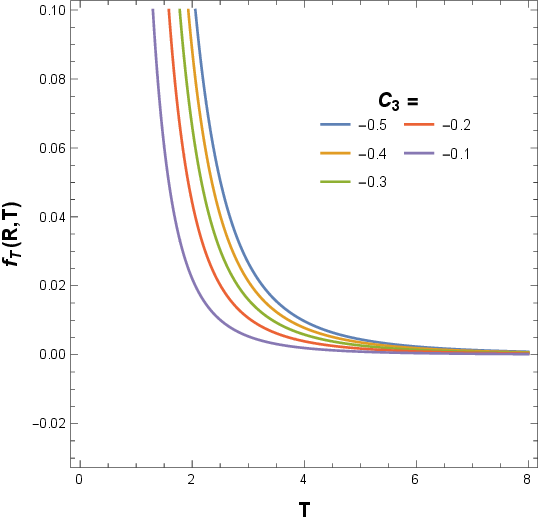}
        \caption{Dark energy dominated epoch}
        \label{12b}
    \end{subfigure}
\end{minipage}
\caption{Variation of $f_T(R,T)$ vs $T$ for for all values of $m$ in (a) matter and (b) dark energy dominated epoch}
\end{figure*}

\begin{figure*}
\begin{minipage}[b]{0.4\linewidth}
\centering
 \begin{subfigure}[b]{0.9\textwidth}
        \includegraphics[width=\textwidth]{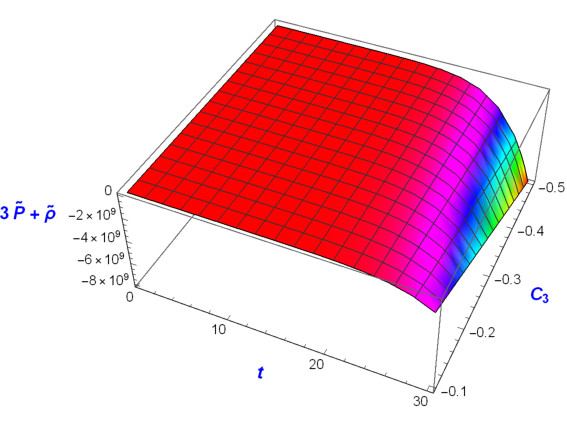}
        \caption{Matter dominated epoch}
        \label{13a}
    \end{subfigure}
\end{minipage}
\hspace{2cm}
\begin{minipage}[b]{0.4\linewidth}
\centering
 \begin{subfigure}[b]{0.9\textwidth}
        \includegraphics[width=\textwidth]{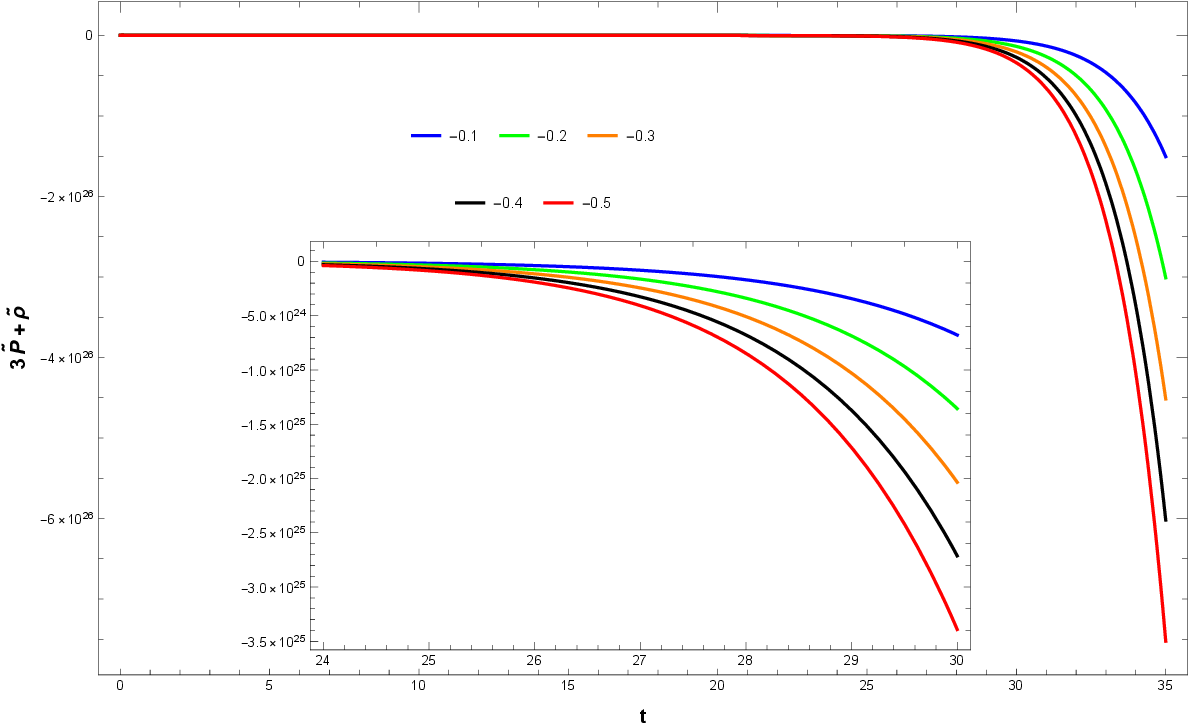}
        \caption{Dark energy dominated epoch}
        \label{13b}
    \end{subfigure}
\end{minipage}
\caption{Variation of ($3\tilde{p}+\tilde{\rho}$) with $t$, fixed $C_2~(or~C'_{2}) = 0, C_1~(or~C'_{1}) = 0.3, m = 5$ for different values of $C_3~(or~C'_{3})$ in (a) matter and (b) dark energy dominated epoch}
\end{figure*}

\begin{figure*}
\begin{minipage}[b]{0.4\linewidth}
\centering
 \begin{subfigure}[b]{0.9\textwidth}
        \includegraphics[width=\textwidth]{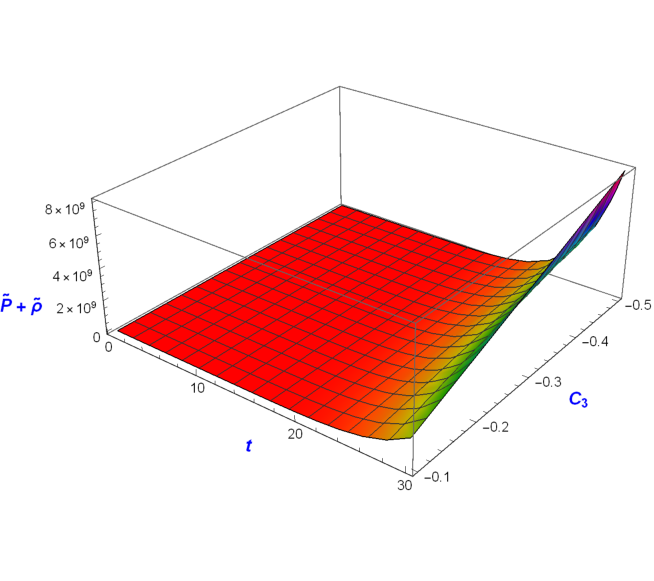}
        \caption{Matter dominated epoch}
        \label{14a}
    \end{subfigure}
\end{minipage}
\hspace{2cm}
\begin{minipage}[b]{0.4\linewidth}
\centering
 \begin{subfigure}[b]{0.9\textwidth}
        \includegraphics[width=\textwidth]{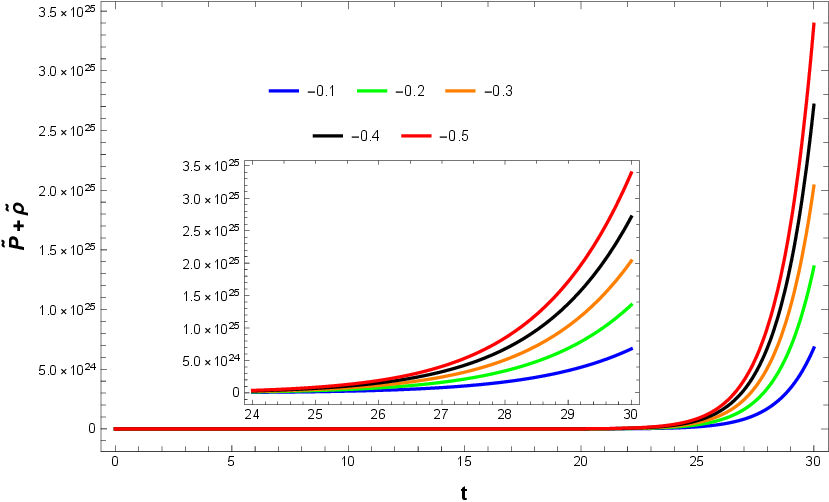}
        \caption{Dark energy dominated epoch}
        \label{14b}
    \end{subfigure}
\end{minipage}
\caption{Variation of ($\tilde{p}+\tilde{\rho}$)  with $t$,
fixed $C_2~(or~C'_{2}) = 0, C_1~or~ (C'_{1})= 0.3,~m = 5$ for different values of $C_3~(or~C'_{3})$ in (a) matter and (b) dark energy dominated epoch}
\end{figure*}

\begin{figure*}
\begin{minipage}[b]{0.4\linewidth}
\centering
 \begin{subfigure}[b]{0.9\textwidth}
        \includegraphics[width=\textwidth]{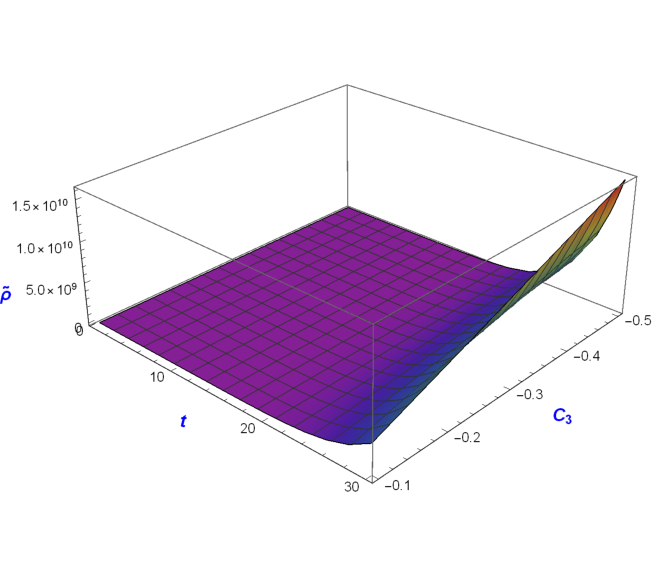}
        \caption{Matter dominated epoch}
        \label{15a}
    \end{subfigure}
\end{minipage}
\hspace{2cm}
\begin{minipage}[b]{0.4\linewidth}
\centering
 \begin{subfigure}[b]{0.9\textwidth}
        \includegraphics[width=\textwidth]{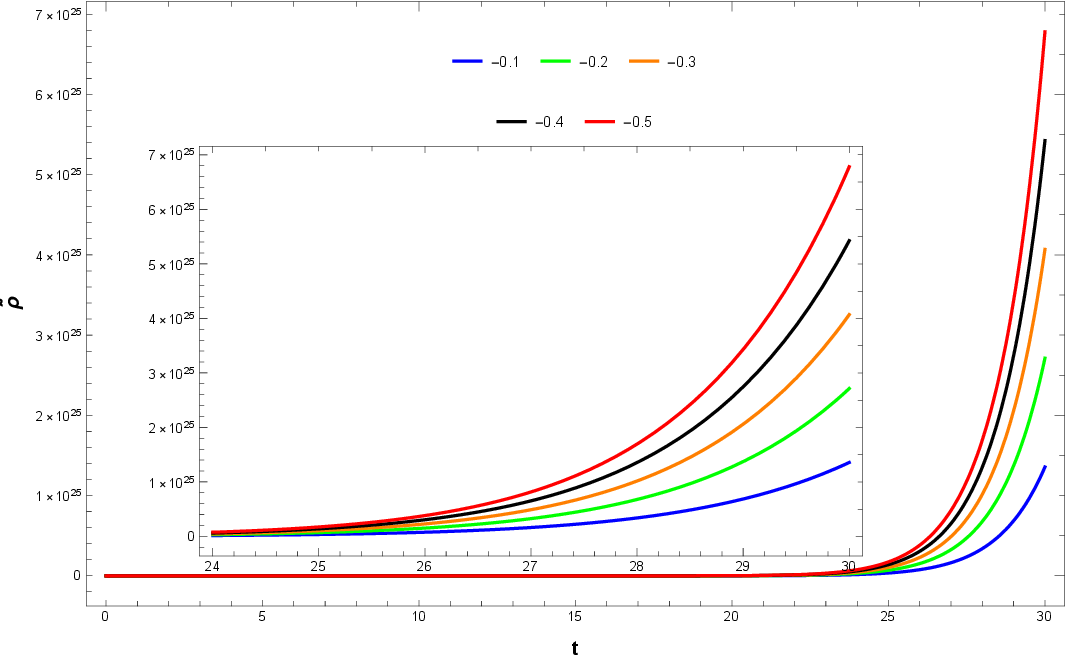}
        \caption{Dark energy dominated epoch}
        \label{15b}
    \end{subfigure}
\end{minipage}
\caption{Variation of $\tilde{\rho}$  with $t$, fixed $C_2 (or~C'_{2}) = 0, C_1 (or~C'_{1})= 0.3, m = 5$ for different values of $C_3~(C'_{3})$
in (a) matter and (b) dark energy dominated epoch}
\end{figure*}

\begin{figure*}
\begin{minipage}[b]{0.4\linewidth}
\centering
 \begin{subfigure}[b]{0.9\textwidth}
        \includegraphics[width=\textwidth]{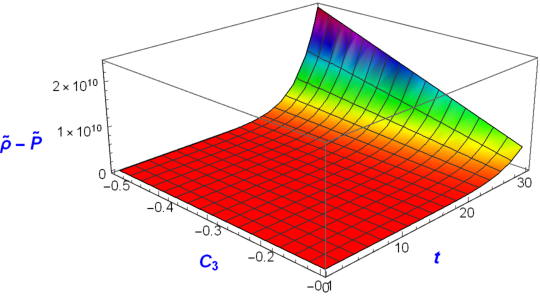}
        \caption{Matter dominated epoch}
        \label{16a}
    \end{subfigure}
\end{minipage}
\hspace{2cm}
\begin{minipage}[b]{0.4\linewidth}
\centering
 \begin{subfigure}[b]{0.9\textwidth}
        \includegraphics[width=\textwidth]{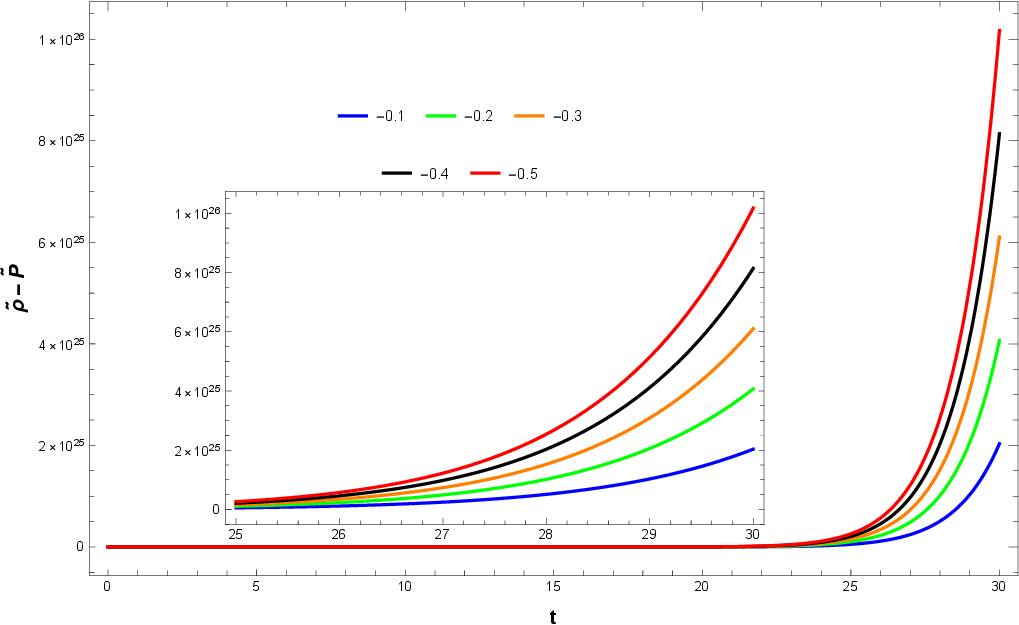}
        \caption{Dark energy dominated epoch}
        \label{16b}
    \end{subfigure}
\end{minipage}
\caption{Variation of $\tilde{\rho}-\tilde{p}$  with $t$,
fixed $C_2 (or~C'_{2}) = 0,~C_1 (or~C'_{1})= 0.3, m = 5$ for different values of $C_3~(C'_{3})$
 in (a) matter and (b) dark energy dominated epoch}
\end{figure*}

Now we are discussing the aspects of viability for our models. From Figs. 6(a) and 6(b) we note that $f_{R}(R,T)$ is always positive for any values of $R$ in both matter-dominated (Fig. 6(a)) and dark energy-dominated epoch (Fig. 6(b)) for $m=1$, respectively. For the same values of $R$, $f_{R}(R,T)$ is slightly greater in the dark energy-dominated epoch, e.g., at $R=1000$, the value of $f_{R}(R,T)$ in the matter-dominated epoch is $4.21129$ for say $C_1=1$ whereas, in the dark energy dominated epoch it is $17.1718$. This means this theory is a little bit more viable for dark energy-dominated epoch for $m=1$. It should be noted that we have used the usual cosmological units for the values of Ricci scalar ($R$) throughout the article.

For $m=5$, it can be seen from Fig. 7(a) that $f_{R}(R,T)$ becomes negative at $R>2000$ for matter-dominated epoch, whereas, from Fig. 7(b), $f_{R}(R,T)$ becomes negative at $R>40,000$ for the dark energy-dominated epoch. i.e., the dark energy-dominated epoch is more viable at higher curvature.

In the case of $m=20$ again $f_{R}(R,T)$ is always positive for both the matter-dominated epoch (Figure 8(a)) and dark energy-dominated epoch (Fig. 8(b)). Note that the nature and value of $f_{R}(R,T)$ are exactly the same in both cases. This means at higher $m$ values, different epoch shows similar kind of variation.

Figures 9(a) and 9(b) show the positive variations of $f_{RR}(R,T)$ with $R$ for $m=1$ in the matter and the dark energy-dominated epochs respectively. Though the variation is similar for both epochs but, in the dark energy-dominated epoch the value of $f_{RR}(R,T)$ is slightly higher than in the matter-dominated epoch, e.g., for say $C_1=5$ at $R=8000$ the value of $f_{RR}(R,T)$ in matter-dominated epoch and dark energy-dominated epoch is $0.0061$ and $0.00838$, respectively.

In Fig. 10(a), it is seen that the variation of $f_{RR}(R,T)$ becomes negative at $R\approx 1500$ suddenly in matter-dominated epoch for $m=5$ whereas, in Fig. 10(b) the value of $f_{RR}(R,T)$ decreases slowly and takes a negative value at $R\approx 20,000$. Again this shows the greater viability in the dark energy-dominated epoch.

Figures 11(a) and 11(b) show the variation of $f_{RR}(R,T)$ for $m=20$ in matter and dark energy-dominated epochs, respectively. The value of $f_{RR}(R,T)$ remains positive up to almost $R=8000$ in the matter-dominated epoch whereas this value remains positive up to $R=15,000$ in the dark-energy-dominated epoch. 

In each scenario, our model is clearly more reasonable for the dark energy-dominated epoch at larger curvature than for the matter-dominated epoch. This is logical because we are working with a scale factor that leads to the universe expanding eternally. The expansion of our universe is more obvious in the era dominated by dark energy. The preceding explanations are also observationally relevant.

As there is no contribution of $m$ in the last term of Eqs. (\ref{44}) and (\ref{52}), we get two single plots for the two different examples. The variation of $f_{T}(R,T)$ says us that it is always positive if we take the constant $C_3$(or$~{C_3}^{\prime})$ is negative, arbitrarily. For different eras, the steepness of the curve just varies as $T$ in these two models has different numbers in their index. But the positivity of $f_{T}(R,T)$ always confirms the stability of these models. Due to the stability of $f_{T}(R,T)$, we choose $C_3$(or$~{C_3}^{\prime})<0$ as discussed earlier. In both the Figs. 12(a) and 12(b), we vary $C_3$(or$~{C_3}^{\prime})$ from $-0.1$ to $-0.5$ and we see that for higher negative values of $C_3$(or$~{C_3}^{\prime})$, the stability in terms of $f_{T}(R,T)$ gives a higher positive value at a fixed $T$. As in the case of the accelerated dark energy-dominated universe, the trace of the energy-momentum tensor varies as $T^{-\frac{5}{2}}$, the plot in Fig. 12(b) becomes condensed than in the case of the matter-dominated era where the contribution in $f(R,T)$ due to the matter part is just $T^{-\frac{1}{2}}$.

\subsection{Features of Energy Conditions}~\label{EC}
In this section, we are going to investigate whether or not the energy conditions applicable to this $f(R, T)$ gravity model are satisfied. We have four energy conditions to check, namely, strong energy condition ({\it SEC}) (Eq. (\ref{30})), null energy condition ({\it NEC}) (Eq. (\ref{31})), weak energy condition ({\it WEC}) (Eq. (\ref{32})) and dominant energy condition ({\it DEC}) (Eq. (\ref{33})). We have various parameters to set, to plot these graphs. We consider the valid range of values of the constants $(C_1$(or$~{C_1}^{\prime})$, $C_2$(or$~{C_2}^{\prime})$, and $C_3$(or$~{C_3}^{\prime})$, as specified in the previous subsection \ref{EC} and use them to test the preservation of energy conditions. For the stability analysis in terms of the positivity of $f_{R}(R,T)$ and $f_{RR}(R,T)$, we've already chosen $C_1$(or $~{C_1}^{\prime})$ to be a positive integer greater than $0$ for our model. And the viability analysis in terms of $f_{T}(R,T)$ tells us that $C_3$(or$~{C_3}^{\prime})$ should always be a negative for this model. So in this section, we take these constants as mentioned above and choose a fixed $C_1$(or$~{C_1}^{\prime})=5$ and note that $C_2$(or$~{C_2}^{\prime})$ has already been chosen to be zero.

We see from Figure 13(a) that SEC violates in the matter-dominated epoch. As the violation of SEC tells us about the expanding nature of our universe, we can say that there exists a signature of expansion in the matter-dominated epoch. Figure 13(b) also says that SEC violates for dark energy-dominated epoch i.e., it takes a negative value, which is obvious for the accelerating present universe. Note that, the negativity is much more in the dark energy case $\mathcal{O}(10^{26})$ than in matter-dominated era $\mathcal{O}(10^{9})$. In both cases, it starts from zero and then goes to a negative value. Now as $\Tilde{\rho}+3\Tilde{p}=0$ is also a condition for holding the SEC, we may infer that at the early time of our universe (lower value of $t$) SEC was satisfied up to some range.

But as time progresses, SEC starts to take the negative value for the matter-dominated era (Figure 13(a)) and takes some forward values of time for the accelerating model of the universe and starts to get violated.
An interesting nature of the violation of SEC in the case of the dark energy-dominated epoch was also found. It is found from Figure 13(b) that if we increase the range of time, the graphs always end up being negative at that value of time, which may be an indication that SEC keeps going to be negative and more negative in future. As a result, we may conclude that the universe expands indefinitely. 

On the other hand, our model is consistent with the other energy conditions for the matter and dark energy-dominated epochs, as shown in Figs. 14 (NEC), 15 (WEC) and 16 (DEC), respectively. It should also be noted that for dark energy-dominated epochs, the pictures are 2-dimensional since higher $y-$axis values cannot produce a 3-dimensional plot.

\section{Discussion and Conclusion}
In the present study we attempted to formulate a geometrical and analytical solution to the $f(R,T)$ gravity model using the Raychaudhuri equation. The motivation for doing this study via the Raychaudhuri equation is as follows: since the Raychaudhuri equation is completely a geometrical equation that deals with the kinematics of flows, the statements of the evolution equations are independent of any sources of Einstein's General Relativity~\cite{Ray}. So at first, the field equation Eq. (\ref{4}) of $f(R,T)$ gravity theory has been rearranged and expressed as $R_{\a\b}$ in Eq. (\ref{9}). Then it is used to determine the gravity-affected term (which is curvature related) of the Raychaudhuri equation Eq. (\ref{10}). In this process, we have taken the line element as a flat FLRW-type and the cosmic fluid behaves like a perfect fluid. Afterwards, from the definition of expansion ($\th$) in a flat FLRW metric, we determine the Raychaudhuri equation for $f(R,T)$ gravity. In the absence of shear and rotation tensor, this equation leads us to the final form (though as a special form) of the Raychaudhuri equation (Eq. (\ref{37})). In the meanwhile, we briefly discussed all the energy conditions in $f(R,T)$ gravity theory with the choice of flat FLRW-type metric and perfect fluid consideration. In the next phase of our work, we choose a power law solution of the scale parameter (Eq. (\ref{38})) for simplicity. By using the form of the scale parameter we have found out the solution of the Raychaudhuri equation for $f(R,T)$ gravity in two epochs, i.e., matter-dominated and dark energy accelerating model of the universe. The form of $f(R,T)$ contains different powers of $m$ ($\a,\b,\d$) and some arbitrary constants ($C_1$(or$~{C_1}^{\prime})$, $C_2$(or$~{C_2}^{\prime})$ and $C_3$(or$~{C_3}^{\prime})$). Now the restriction upon $m$ is that $m$ can have any value greater than $0.5$ which  may be a fraction as well as an integer. So, eventually we decide to check the variation of the different powers of $R$ with $m$ for $m=1,~5$ and $20$. It is to be noted that we are dealing with an eternally accelerated power law expansion of the model and from $m=0.5$, $\b$ takes a negative value with increasing $m$ leading to the negative curvature effect. Therefore we discard the $R^\b$ term from the general solution of $f(R,T)$ model by choosing the arbitrary constant $C_2$ (or$~{C_2}^{\prime})$ to be $0$. 

Next, we attempt to analyse our model whether it is viable or not. Therefore the viability conditions are discussed in subsection A of section-VI. To do this we plot the variation of $f_{R}(R,T)$ with respect to $R$. Here one can note from Eq. (\ref{7}) that $f_{R}(R,T)$ should remain always positive to make the Einstein tensor ($G_{\a\b}$) well-defined and the effective gravitational constant ghost free. 

From Figs. 6 and 8, it is clear that $f_{R}$ remains always positive satisfying the first condition of viability for both the matter and dark energy-dominated era for the chosen values, e.g. $m=1$ and $m=20$, respectively. Interestingly, for $m=1$ the viability w.r.t. $f_{R}(R,T)$ is greater in the dark energy-dominated epoch than in the matter-dominated epoch but there is no difference in the value of $f_{R}(R,T)$ in the two epochs for $m=20$. From Fig. 7 with $m=5$, the dark energy-dominated epoch is more viable than matter dominated epoch at higher curvature as the positivity of $f_{R}(R,T)$ holds up to a larger value of $R$ in case of dark energy-dominated epoch ($40,000$) than in the matter-dominated epoch ($2000$). 

In this process, we choose the arbitrary constants as mentioned in section VI-A and the value of $\frac{\rho_0}{C^3}$ (for the matter-dominated era) and $\frac{\rho_{0}}{C^{3/2}}$ (for the dark energy-dominated era) has been taken as $1$. The stability condition for this theory implies that $f_{RR}$ must be greater than $0$, which is satisfied in both epochs for $m=1$ forever. However, for the case of $m=5$ and $20$, the viability breaks at different values of $R$ for two different epochs. This may happen due to choice of the scale factor which follows a power law expansion. Using the choices of the arbitrary constants, mentioned in section VI-A, we check all the energy conditions in this theory and it is seen that all the energy conditions are well-satisfied except the strong energy condition. 

Since we obtain a solution to the Raychaudhuri equation in $f(R,T)$ theory on the basis of power law expansion of the cosmic scale factor, it is inevitable that we would obtain the violation of SEC. This violation is more prominent in the accelerated dark energy-dominated universe rather than the matter-dominated one. It is arguable that this violation of SEC is observationally relevant.

From the viability analysis, we can say from a completely mathematical point of view that in comparison with the works~\cite{Shibendu, Shabani} the $f(R,T)$ gravity model is much more stable at high curvature than $f(R)$ gravity. Basically from the viability analysis and the study of energy conditions, one may note that $C_1$ (or$~{C_1}^{\prime})$ is always a positive constant which is chosen by an integer. On the other hand, $C_2$ (or$~{C_2}^{\prime})=0$ (and also $C_3$ (or$~{C_3}^{\prime})$) is a negative constant. Except for these particular choices of arbitrary constants, all the energy conditions are not valid in both the matter and dark energy-dominated era, i.e., we can argue that our theory is self-consistent.

Now, we would like to address here the Raychaudhuri equation in $f(R,T)$ gravity as a harmonic oscillator equation (\ref{36}) with variable frequency. Interestingly, Eq. (\ref{36}) can also be used to analyze the expansion equation. It is worth noting that the expansion $\th$ is nothing else than the rate of change of the cross-sectional area orthogonal to the bundle of geodesics~\cite{Kar}. The harmonic oscillator equation (\ref{36}) specifies the condition for convergence as $(R_{\a\b} v^{\a} v^{\b}+2\sg^2-2\o^2)\geq 0$, where $R_{\a\b}$ in the $f(R,T)$ theory is represented by Eq. (\ref{9}). For such hypersurface orthogonal congruences (zero rotation), the convergence condition is then very straightforward: $R_{\a\b} v^{\a} v^{\b}\geq 0$ which leads to geodesic focusing~\cite{Kar}. The SEC (\ref{16}) can be derived from the standard Einstein field equations by rewriting the Ricci tensor in terms of the energy-momentum tensor. This implies that if the SEC is to be obeyed, geodesic focusing encodes the straightforward proposition that geodesics must eventually be drawn towards one another if the matter is attractive. However, in the present model, the SEC in $f(R,T)$ gravity is violated in both the matter-domination and dark energy-domination periods implying that the universe is expanding rather than contracting. This is fundamentally valid based on the present-day observational results. On the other hand, the Raychaudhuri equation in harmonic oscillator form (\ref{36}) can also be applied to study the quantum behavior of the cosmos. The quantum behavior of the Raychaudhuri equation is beyond the scope of this article.

As a final comment it is important to mention that the use of the Friedmann equations is not applicable in the present study as has been discussed in the earlier part of the manuscript. Instead, for the time being, our focus is on the Raychaudhuri equation (\ref{37}) alone. However, in future, we can utilize the modified Friedmann equations to analyze other cosmological instances.

\vspace{0.5in}

{\bf Acknowledgements:}
AP and GM acknowledge the DSTB, Government of West Bengal, India for support through the Grants No.: 322(Sanc.)/ST/P/S\&T/16G-3/2018 dated 06.03.2019. SR is thankful to the Inter-University Centre for Astronomy and Astrophysics (IUCAA), Pune, India for providing Visiting Associateship under which a part of this work was carried out who also gratefully acknowledges the facilities under ICARD, Pune at CCASS, GLA University, Mathura. Additionally, the authors express their gratitude to the referees for providing insightful comments to enhance
the quality of the work.

\vspace{0.5in}

{\bf DATA AVAILABILITY:}
There are no assoiated data with this artile as
such no new data were generated or analyzed in support of this research.

\end{document}